\documentclass[prd,preprint,double-spaced,superscriptaddress,showpacs]{revtex4}
\begin{document}
\title{Obtaining the Time Evolution for Spherically Symmetric Lema\^{i}tre-Tolman-Bondi Models Given Data on Our Past Light Cone} 

\date{\today}

\author{M.E. Ara\'{u}jo}
\affiliation{ Departamento de F\'{\i}sica-Matem\'{a}tica, Instituto de F\'{\i}sica,\\
    Universidade Federal do Rio de Janeiro,\\
            21.945-970, Rio de Janeiro, R.J., Brazil}
\author{W.R. Stoeger}
\affiliation{ Vatican Observatory Research Group \\
      Steward Observatory, University of Arizona, \\
      Tucson, AZ 85721, USA}

\begin{abstract}

A rigorous demonstration that given appropriate data on our past light cone leads to the determination of the metric functions and all their time derivatives on our past light cone is presented, thus showing how to evolve the solution we obtain from data on the light cone off it in a well-defined and straightforward way. It also automatically gives a procedure for constructing the solution for all spherically symmetric,
inhomogeneous cosmological Lema\^{i}tre-Tolman-Bondi  models in observational coordinates
as a Taylor series in time of however many terms we need. 
Our procedure takes into account the essential data giving the maximum of the observer
area (angular-diameter) distance, and the redshift $z_{max}$ at which that occurs. This enables the determination of the vacuum-energy density $\mu_{\Lambda}$, which would otherwise remain 
undetermined. \\

 \end{abstract}


\maketitle

\section{Introduction}

Since the pioneer work of Hubble in the twenties, it has been well known that the time span of cosmological observations is very small compared to the age of our Universe. Consequently,
all we can aim to obtain on a
cosmological time scale is data on one single light cone (our past light cone), which, for all practical purposes, can be considered the same as Hubble's. That means that we do
not have the time variation of any cosmological data to start with.
For this reason, observational cosmological modeling in which one wants to find the metric based only on observed quantities has been very much restricted to
modeling our past light cone. That has been a handicap researchers in this
field have been trying to overcome for some time -- with very limited success -- and a key issue for the future development of observational cosmology.    

Given this lack of knowledge of the time variations of cosmological data and assuming that the Universe is spherically symmetric around us, the only way of finding the time variation of a cosmological quantity is to derive a Taylor series. But that demands knowledge of  the quantity itself and all its time derivatives on our past light cone, if we use observational coordinates.

In this paper, we present a rigorous demonstration that, given appropriate data on our past light cone, we
can find the metric functions and all their time derivatives on our past light 
cone, thus showing how to evolve the solution off it into the past or into the future. Besides being important in its own right,
it also automatically gives us a procedure for constructing the solution as a Taylor series in time of however many terms we need. Moreover, our procedure fully takes into
account an important piece of data, the maximum of the observer area distance, and the redshift at which it occurs. Without these extra observables, we do not have enough
independent data to determine the model -- in particular to determine the cosmological constant  $\Lambda$. An intermediate and necessary step for doing that is to show how an
important result originally obtained by Hellaby  \cite{Hellaby} in the Lema\^{i}te-Tolman-Bondi (LTB) $3+1$ formalism also follows from the observational cosmology field equations.

In several papers Ara\'ujo and Stoeger \cite{AS}, Ara\'ujo {\it et al.} and \cite{AABFS},  
Ara\'ujo, Roveda and Stoeger \cite{ASR} demonstrated how to solve exactly the
spherically symmetric (SS) Einstein field equations  for dust in observational 
coordinates without assuming Friedmann-Lema\^{i}tre-Rebertson-Walker (FLRW) and with cosmological data representing galaxy
redshifts, observer area distances and galaxy number counts as
functions of redshift. All these papers assumed that $\Lambda = 0$.  
In a recent paper,  Ara\'ujo {\it et al.} \cite{ASAB} we demonstrated how this program may be 
carried out when $\Lambda \neq 0$. These data are given, not on a space-like
surface of constant time, but rather on our past light cone $C^-(p_0)$, which is centered at 
our observational position $p_0$ ``here and now'' on our world line $ {\cal C}$.  These results
demonstrate how cosmologically relevant astronomical data can be used to determine the space-time 
structure of the universe -- the cosmological model which best fits it. This has been the aim
of a series of papers going back to the Physics Reports paper by Ellis
{\it et al.} \cite{Ellis et al}. The motivation and history of this `` observational
cosmology (OC) program''  is summarized in Ara\'ujo and Stoeger \cite{AS}.

In all these papers, we have shown that if  the redshift,
observer area-distance, and number-count data can be fit to FLRW functional
forms (these are very special forms the data must take, if the universe
is FLRW) then  such data determines a {\it bona fide} FLRW
universe -- which is not {\it a priori} obvious.

 For that purpose, we have used
an integration scheme that has been improved over the years, but still has had an element 
of uncertainty about it, in the sense that we did not have a precise mathematical way of
determining the time evolution of one of the metric functions $A(w,y)$ (see below)
which is fundamental for the solution of the problem. That was due to the fact that
apparently we have the gauge freedom of choosing $A(w,0)$ in any way we like.
However, our experience, acquired in treating all these cases, has shown that,
unless $A(w,0)$ has a very specific functional dependency on $w$, the solution thus
obtained would not satisfy the central conditions. Unless the central conditions 
are satisfied, we are not guaranteed  that the null surface on which we assume we have 
the data is a past light cone of our world line. More specifically, we found that using the 
apparent gauge freedom of choosing $A(w,0)$ at will leads to a formal solution of the field
equations which does not necessarily satisfy the necessary boundary conditions,
namely, the central conditions.

 Thus, it has become clear that in fact we do not have a choice of choosing
$A(w,0)$ since, in order to find a solution that also satisfies the central conditions we have, 
in a sense, to work backwards from the central conditions trying to figure out among all ``possible``
choices of $A(w,0)$ the one that leads to a solution satisfying the central conditions. We were successful
in applying this procedure to various FLRW models but it was not satisfactory, because it is not
an algorithmic procedure and depends on one's ability to spot the correct functional 
dependency that leads to the desired solution.

 Our next step was to try  to understand what causes
this loss of gauge freedom of choosing $A(w,0)$ and in Ara\'ujo, Roveda and Stoeger \cite{ASR}  and
Ara\'ujo {\it et al.} \cite{ASAB} we argued that the fulfillment of the following conditions:
{\it

(1) $A(w_0,y)$ is determined  by the data and the central conditions;

(2) The coordinate $y$ is choosen to be a comoving radial coordinate;

(3) The central conditions }(\ref{cent})

\noindent {removes the freedom of rescaling the time coordinate $w$ and
completely determines $A(w,y)$.}  That is indeed a correct set of assumptions
for proving that conjecture, but we were not able then to present a rigorous mathematical 
proof of the result. Here, as we shall demonstrate, we have been successful in proving that the
more restricted set of assumptions, namely (1) and (3), leads  to the desired result for all inhomogeneous
spherically symmetric LTB models, thus filling this important gap in observational cosmology modeling.

Recently, Hellaby and Alfadeel \cite{helalf}, after rederiving the key OC equations
in a somewhat different way and emphasizing the free functions which must
be determined for their solution, developed an integration scheme for determining the time 
evolution of $A(w,y)$ -- as well as the other metric functions $B(w,y)$ and $C(w, y)$ -- but only indirectly, 
by first determining the time function $t(w,y)$ along all the matter world lines, and then finding
$A(w, y)$ from that $A = t_w$ (see Eq. (\ref{tw}) below). This is because they could not find an
evolution equation for $A_w$, as they themselves emphasize. Here we demonstrate how to
do that, which leads to a much more direct and streamlined integration scheme
-- without having to resort to integrating over the $3+1$, $(t,y)$ domain first.

Furthermore, as mentioned earlier, we show how the data (especially the redshift and the observer-area -- 
or angular-diameter -- distance data) in conjunction with the central conditions eliminate the apparent
gauge freedom and completely determine the solution both on our past light cone, and off it into the past.

The primary aim of OC program is to strengthen the connections
between astronomical observations and cosmological theory. We do this
by allowing observational data to determine the geometry of spacetime
as much as possible, {\it without} relying on {\it a priori}
assumptions more than is necessary or justified. Basically, we want to
find out not only how far our observable universe is from being
isotropic and spatially homogeneous (that, is describable by an
FLRW cosmological model) on various length scales, but also to give a 
dynamic account of those deviations (Stoeger {\it et al} \cite{OC III}).

By using observational coordinates, we can thus formulate
Einstein's equations in a way which reflects both the geodesic flow of the
cosmological fluid and the null geometry of $C^-(p_0)$, along which 
practically all of our information about the distant reaches of our 
universe comes to us -- in photons. In this formulation the field equations
split naturally into two sets, as can be easily seen: a set of equations
which can be solved on $C^-(p_0)$, that is on our past light cone, specified
by $w = w_0$, where $w$ is the observational time coordinate; and a second
set which evolves these solutions off $C^-(p_0)$ to other light cones into
the past or into the future. Solution to the first set is directly determined 
from the data, and those solutions constitute the ``initial conditions'' for the
solution of the second set.

In this paper, for completeness, we review some aspects of the problem
of determining the solution of the exact spherically symmetric
Einstein equations for dust in observational coordinates. 
We refer the reader to Ellis {\it et al} \cite{Ellis et al}, Kristian and Sachs
\cite{KS}, Ara\'ujo and Stoeger \cite{AS} and references therein for a complete
account of the philosophy and the foundations of the OC approach leading to
the integration of Einstein field equations in observational coordinates.

In the next section we define observational coordinates, write the
general spherically symmetric metric using them and present the very
important central conditions for the metric variables. Section \ref{sec:obspar}
summarizes the basic observational parameters we shall be using and
presents several key relationships among the metric variables.
Section \ref{sec:fieqs} presents the full set of field equations for the spherically
symmetric case, with dust and with $\Lambda \neq 0$. In Section \ref{sec:nonflat}, we
present a general integration scheme for all inhomogeneous spherically symmetric LTB models
and in section \ref{sec:concl} we briefly discuss our conclusions.

\section{ The Spherically Symmetric Metric in Observational Coordinates}

We are using observational coordinates (which were first suggested by
Temple \cite{Temple}). As described by Ellis {\it el al}  \cite{Ellis et al}  the observational
coordinates $x^i=\{w,y,\theta ,\phi \}$ are centered on the observer's
world line $ {\cal C}$ and defined in the following way: \\

\noindent
(i) $w$ is constant on each past light cone along $ {\cal C}$, with $u^a
\partial _a w > 0$ along $ {\cal C}$, where $u^a$ is the 4-velocity of matter
($u^au_a=-1$). In other words, each $w = constant$ specifies a past
light cone along $ {\cal C}$. Our past light cone is designated as $w =
w_0$. \\

\noindent
(ii) $y$ is the null radial coordinate. It measures distance down the null
geodesics -- with affine parameter $\nu$ -- generating each past light cone
centered on $ {\cal C}$. $y = 0$ on $ {\cal C}$ and $dy/d\nu > 0$ on each null cone -- so
that $y$ increases as one moves down a past light cone away from $ {\cal C}$. \\

\noindent
(iii) $\theta$ and $\phi$ are the latitude and longitude of 
observation, respectively -- spherical coordinates based on a
parallelly propagated orthonormal tetrad along $ {\cal C}$, and defined away
from $ {\cal C}$ by $k^a \partial _a \theta = k^a \partial _a \phi = 0$, where
$k^a$ is the past-directed wave vector of photons ($k^ak_a=0$). \\

\noindent
There are certain freedoms in the specification of these observational
coordinates. In $w$ there is the remaining freedom to specify $w$ along our
world line $ {\cal C}$. Once specified there it is fixed for all other world lines.
There is considerable freedom in the choice of $y$ -- there are a large
variety of possible choices for this coordinate -- the affine parameter, $z$%
, the area distance $C(w,y)$ itself. We normally choose $y$ to be comoving
with the fluid, that is $u^a\partial _ay=0$. Once we have made this choice,
there is still a little bit of freedom left in $y$, which we shall use below.
The remaining freedom in the $\theta $ and $\phi $ coordinates is a rigid
rotation at $one$ point on $ {\cal C}$. 

In observational coordinates the Spherically Symmetric metric takes
the general form:
\begin{equation}
ds^2=-A(w,y)^2dw^2+2A(w,y)B(w,y)dwdy+C(w,y)^2d\Omega ^2,  \label{oc}
\end{equation}
where we assume that $y$ is comoving with the fluid, so that the
fluid 4-velocity is $u^a=A^{-1}\delta _w^a$.

The remaining coordinate freedoms which preserves the observational form of
the metric is a scaling of $w$ and of $y$:
\begin{equation}
w\rightarrow \tilde{w}=\tilde{w}(w)~,~~y\rightarrow\tilde{y}= \tilde{y}%
(y)~~~~\left({\frac{d\tilde{w}}{dw}}\neq 0 \neq {\frac{d\tilde{y} }{dy}}%
\right).  \label{wy}
\end{equation}

The first, as we mentioned above, corresponds to a freedom to choose $w$ as
any time parameter we wish along $ {\cal C}$, along our world line at $y=0$. This is
usually effected by choosing $A(w,0)$. As we have briefly indicated already, this
freedom disappears when we apply the central conditions (see below). The second corresponds to the freedom
to choose $y$ as any null distance parameter on an initial light cone --
typically our light cone at $w=w_0$. Then that choice is effectively dragged
onto other light cones by the fluid flow. $y$ is comoving with the fluid
4-velocity, as we have already indicated. We shall use this freedom to
choose $y$ by setting:
\begin{equation}
A(w_0, y) = B(w_0, y).  \label{ab}
\end{equation}
We should carefully note here that setting $A(w, y) = B(w, y)$ off our
past light cone $w = w_0$ is too restrictive.

In general, these freedoms in $w$ and $y$ imply the metric scalings:
\begin{equation}
A\rightarrow\tilde{A}={\frac{dw}{d\tilde{w}}}A~,~~ B\rightarrow\tilde{B}={%
\frac{dy}{d\tilde{y}}}B.  \label{4scale}
\end{equation}

It is important to specify the central conditions for the metric variables $%
A(w, y)$, $B(w, y)$ and $C(w, y)$ in Eq. (\ref{oc}) -- that is, their proper
behavior as they approach $y = 0$. These are:
\begin{eqnarray}
{\rm as}\;\;y\rightarrow 0:\;\;\; &&A(w,y)\rightarrow A(w,0)\neq 0, 
\nonumber \\
&&B(w,y)\rightarrow B(w,0)\neq 0,  \nonumber \\
&&C(w,y)\rightarrow B(w,0)y = 0,  \label{cent} \\
&&C_y(w,y)\rightarrow B(w,0).  \nonumber
\end{eqnarray}
These important conditions insure that ${\cal C}$, our world line, is
regular -- so that all functions on it our bounded, and that the
spheres ($w$, $y =$ constant) go smoothly to ${\cal C}$ as $y \rightarrow 0$.
They also insure that the null surfaces $w =$ constant  are past light cones
of observers on ${\cal C}$ (See reference  \cite{Ellis et al} , especially section 3.2, p.
326, and Appendix A for details).

\section{\label{sec:obspar}The Basic Observational Quantities}

The basic observable quantities on $ {\cal C}$ are the following: \\

(i) Redshift. The redshift $z$ at time $w_0$ on $ {\cal C}$ for a comoving source a
null radial distance $y$ down $C^{-}(p_0)$ is given by
\begin{equation}
1+z={\frac{A(w_0,0)}{A(w_0,y)}}.  \label{z}
\end{equation}
This is just the observed redshift, which is directly determined by source
spectra, once they are corrected for the Doppler shift due to local motions.
It is consistent with and complements  the first of central conditions in
Eq. (\ref{cent}).
\\

(ii) Observer Area Distance. The observer area distance, often written as $%
r_0$, measured at time $w_0$ on $ {\cal C}$ for a source at a null radial distance $y$
is simply given by
\begin{equation}
r_0=C(w_0,y),
\end{equation}
provided the central condition (\ref{cent}), determining the relation
between $C(w,y)$ and $B(w,y)$ for small values of $y$, holds. This quantity
is also measurable as the luminosity distance $d_L$ because of the reciprocity
theorem of Etherington \cite{Etherington33} (see also Ellis \cite{Ellis 1971}),\\ 

\begin{equation}
d_L = (1+z)^2 C(w_0, y).   \label{recth} \\
\end{equation}

(iii) The Maximum of Observer Area Distance. Generally speaking, $C(w_0, y)$
reaches a maximum $C_{max}$ for a relatively small redshift $z_{max}$ (Hellaby
 \cite{Hellaby}; see also Ellis and Tivon  \cite{ET} and Ara\'{u}jo and Stoeger  \cite{ASII}). At
$C_{max}$, of course, we have 

\begin{equation}
\frac{d C(w_0, z)}{d z} = \frac{d C(w_0, y)}{d y} = 0,\\
\end{equation}
further conditioned by
\begin{equation}
\frac{d^2 C(w_0, z)}{d z^2} < 0.\\
\end{equation}
Furthermore, of course, as we shall
review below,  with the solution of the null Raychaudhuri equation (Eq. (\ref{nr})
below), the data set will give us $y = y(z)$, from which we shall be
able to find $y_{max} = y_{max}(z_{max})$.  These $C_{max}$ and $z_{max}$
data provide additional independent information about the cosmology. Without
$C_{max}$ and $z_{max}$ we cannot constrain the value of $\Lambda$. 
 
(iv) Galaxy Number Counts. The number of galaxies counted by a central
observer out to a null radial distance $y$ is given by
\begin{equation}
N(y)=4\pi\int_0^y \mu(w_0,\tilde{y})m^{-1}B(w_0,\tilde{y})C(w_0,\tilde{y})^2
d\tilde{y},  \label{N}
\end{equation}
where $\mu$ is the mass-energy density and $m$ is the average galaxy mass.
Then the total energy density can be written as
\begin{equation}
\mu(w_0,y) = m\;n(w_0,y) = M_0(z)\;{\frac{dz}{dy}}\;{\frac{1}{B(w_0,y)}},
\label{mudef}
\end{equation}
where $n(w_0, y)$ is the number density of sources at $(w_0, y)$, and where
\begin{equation}
M_0 \equiv {\frac{m}{J}}\;{\frac{1}{d\Omega}}\;{\frac{1}{r_0^2}}\;{\frac{dN}{dz}}. \label{m0def}
\end{equation}
Here $d \Omega$ is the solid angle over which sources are counted, and
$J$ is the completeness of the galaxy count, that is, the fraction of
sources in the volume that are counted is $J$. The effects of dark
matter in biasing the galactic distribution may be incorporated via $m$ and/or
$J$ . In particular, strong biasing is needed if the number counts have
a fractal behaviour on local scales (Humphreys {\it et al} \cite{hmm}). In order
to effectively use number counts to constrain our cosmology, we shall also
need an adequate model of galaxy evolution. We shall not discuss this
important issue in this paper. But, fundamentally, it would give us
an expression for $m = m(z)$ in Eqs. (\ref{mudef}) and (\ref{m0def}) above.

There are a number of other important quantities which we catalogue here for
completeness and for later reference. 

First, there are the two fundamental four-vectors in the problem, the fluid
four-velocity $u^a$ and the null vector $k^a$, which points down the
generators of past light cones. These are given in terms of the metric
variables as
\begin{equation}
u^a = A^{-1}\delta^a{}_w ~,~~ k^a = (AB)^{-1}\delta^a{}_y.  \label{uk}
\end{equation}

Then, the rate of expansion of the dust fluid is $3 H = \nabla_a u^a$, so that,
from the metric (1) we have:
\begin{equation}
H={\frac{1}{3A}}\left({\frac{\dot{B}}{B}}+2{\frac{\dot{C}}{C}}\right),
\label{h}
\end{equation}
where a ``dot'' indicates $\partial/\partial w$ and a ``prime'' indicates $%
\partial/\partial y$, which will be used later. For the central observer $H$
is precisely the Hubble expansion rate. In the homogeneous (FLRW) case, $H$ is
constant at each instant of time t. But in the general inhomogeneous case, $%
H $ varies with radial distance from $y = 0$ on $t = t_0$. From our central
conditions above (3), we find that the central behavior of $H$ is given by
\begin{equation}
{\rm as}\;\;y\rightarrow 0:\;\;\;H(w,y)\rightarrow {\frac{1}{A(w,0)}}{\frac{%
\dot{B}(w,0)}{B(w,0)}}=H(w,0).  \label{hcent}
\end{equation}
At any given instant $w = w_0$ along $y = 0$, this expression is just the
Hubble constant $H_0 \equiv H(w_0, 0) = A_0^{-1} B_0^{-1}(\dot{B})_0$ as
measured by the central observer. In the above we have also written $A_0 \equiv A(w_0, 0)$
and  $B_0 \equiv B(w_0, 0)$.

Finally, from the normalization condition for the fluid four-velocity, we
can immediately see that it can be given (in covariant vector form) as the
gradient of the proper time $t$ along the matter world lines: $u_a=-t,_a$.
It is also given by (\ref{oc}) and (\ref{uk}) as
\begin{equation}
u_a=g_{ab}u^b=-Aw_{,a}+By_{,a}.
\end{equation}
Comparing these two forms implies
\begin{equation}
dt=Adw-Bdy~~\Leftrightarrow~~A=t_w~,~~ B=-t_y,  \label{tw}
\end{equation}
which shows that the surfaces of simultaneity for the observer are given in
observational coordinates by $A dw = B dy$. The integrability condition of
Eq. (\ref{tw}) is simply then
\begin{equation}
A^{\prime}+\dot{B}=0.  \label{coneq}
\end{equation}

This turns out precisely to be the momentum conservation equation, which is
a key equation in the system and essential to finding a solution. 

\section{\label{sec:fieqs}The Spherically Symmetric Field Equations in 
Observational Coordinates}

Using the fluid-ray tetrad formulation of the Einstein's equations 
developed by Maartens \cite{m} and Stoeger {\it et al} \cite{fluid ray}, 
one obtains the Spherically Symmetric field equations in observational 
coordinates with $\Lambda \neq 0$ (see Stoeger {\it et al} \cite{OC III} for a
detailed derivation). Besides the momentum conservation Eq. (\ref{coneq}), they are
as follows:

A set of two very simple fluid-ray tetrad time-derivative equations:

\begin{eqnarray}
\dot{\mu}_m &=& -2{\mu_m} \left(\frac{\dot B}{2B} + \frac{\dot{C}}{C} \right), \label{mueqr} \\
\dot{\omega} &=& -3 \frac{\dot C}{C} \biggl(\omega + \frac{\mu_{\Lambda}}{6} \biggr), \label{omegaeqr}
\end{eqnarray}
where $\mu_m$ again is the relativistic mass-energy density of the dust, 
including dark matter, and 
\[\omega(w,y)\equiv -{\frac{1}{2C^2}} + {\frac{
\dot{C}}{{AC}}}{\frac{C^{\prime}}{{BC}}} + {\frac{1}{2}} \biggl({\frac{
C^{\prime}} {BC}}\biggr)^2,\]
is a quantity closely related to $\mu_{m_0}(y)\equiv \mu_m(w_0,y)$ (see Eq.\ref{omegdef}) below).

Equations (\ref{mueqr}) and (\ref{omegaeqr}) can be quickly integrated to give:
\begin{widetext}
\begin{equation}
\mu_m(w,y)=\mu_{m_0}(y)\;{\frac {B(w_0,y)} {B(w,y)}}\; \frac {C^{2}(w_0,y)} {C^{2}(w,y)}; 
\end{equation}

\begin{equation}
\omega(w,y)=\biggl(\omega_0(y) + {\frac {\mu_{\Lambda}} {6} }\biggr) {\frac{C^3(w_0, y)}{C^3(w,y)}}
- {\frac {\mu_{\Lambda}} {6}} = {-{\frac{1}{{2C^2}}}+{\frac{
\dot{C}}{{AC}}}{\frac{C^{\prime}}{{BC}}}+{\frac{1}{2}}\biggl({\frac{
C^{\prime}}{{BC}}}\biggr)^2}, \label{omega} 
\end{equation}
\end{widetext}
where $\omega_0(y)\equiv \omega(w_0,y)$ and the last equality in (\ref{omega}) follows from the definition of $\omega$ given above. In deriving and solving these equations, and those below, we have used the
typical $\Lambda$ equation of state, $p_{\Lambda} = - \mu_{\Lambda},$
where $p_{\Lambda}$ and $\mu_{\Lambda} \equiv \frac{\Lambda}{8 \pi G}$ are the pressure and the 
energy density due to the cosmological constant. Both $\omega_0$ and
$\mu_0$ are specified by data on our past light cone, as we shall show.
$\mu_{\Lambda}$ will eventually be determined from the measurement of $C_{max}$
and $z_{max}$. 

The fluid-ray tetrad radial equations are:
\begin{widetext}
\begin{eqnarray}
&{\frac{C^{\prime\prime}}{C}} = {\frac{C^{\prime}}{C}}{\biggl({\frac{
A^{\prime}}{A}} +{\frac{B^{\prime}}{B}}\biggr)} - {\frac{1}{2}}B^2\mu_m;
\label{nr} \\
&\biggl[ \bigl(\omega_0(y) + {\frac {\mu_{\Lambda}} {6} }\bigr)C^3(w_0, y)\biggr]^{\prime} = -{\frac{1}{2}}\mu_{m_0}\;
{B(w_0,y)}\;{C^{2}(w_0,y)}\;{\biggl({\frac{\dot {C}}{A}} +
 {\frac{C^{\prime}}{B}}\biggr)};  \label{omegap} \\
&{\frac{{\dot {C}}^{\prime}}{C}} = {\frac{{\dot B}}{B}}{\frac{C^{\prime}}{C}
} - \biggl(\omega + {\frac {\mu_{\Lambda}} {2}}\biggr)\; A\;B.  \label{prdot}
\end{eqnarray}
\end{widetext}
The remaining ``independent'' time-derivative equations given by the
fluid-ray tetrad formulation are:

\begin{eqnarray}
&&{\frac{{\ddot{C}}}C}={\frac{{\dot{C}}}C}{\frac{{\dot{A}}}A}+ \biggl(\omega + 
{\frac {\mu_{\Lambda}} {2}}\biggr) \;A^2;
\label{Cdd} \\
&&{\frac{{\ddot{B}}}B}={\frac{{\dot{B}}}B}{\frac{{\dot{A}}}A}-2\omega \;A^2-{%
\frac 12}\mu_{m} \;A^2 .  \label{bdd}
\end{eqnarray}
From Eq. (\ref{omegap}) we see that there is a naturally defined
``potential'' (see Stoeger {\it et al} \cite{OC III}) depending only on the radial
null coordinate $y$ -- since the left-hand-side depends only on $y$, the
right-hand-side can only depend on $y$:

\begin{equation}
F(y)\equiv {\frac{\dot{C}}A}+{\frac{C^{\prime }}B},  \label{f1}
\end{equation} 
Thus, from Eq.  (\ref{omegap}) itself
\begin{widetext}
\begin{equation}
\omega_0(y)= - {\frac {\mu_{\Lambda}} {6} }- {\frac{1}{2 C^3(w_0, y)}}\;\int{\mu_{m_0}(y)\;
{B(w_0,y)}\;{C^{2}(w_0,y)}\;F(y)\;dy}.  \label{omegdef}
\end{equation}
Connected with this relationship is Eq. (\ref{omega}), which we rewrite
as 
\begin{equation}
{\frac{{\dot C}}{C}}{\frac{C^{\prime}}{C}}+{\frac{A}{2B}}{\frac{{C^{\prime}}%
^2}{C^2}}-{\frac{AB}{2C^2}} = {\frac{AB}{C^3}\biggl[C_0^3\bigl(\omega_0 + \frac{\mu_{\Lambda}}{6}\bigr) - \frac{\mu_{\Lambda}}{6}{C^3}\biggr]},  \label{f2}
\end{equation}
where $C_0 \equiv C(w_0, y)$.

We can now proceed to recover a simple but very important observational relationship which 
will enable us to determine $\mu_{\Lambda}$. We begin by differentiating Eq.  (\ref{f1}) with
respect to $w$. This gives

\begin{equation}
\dot {C}^{\prime} = \dot B \Biggl( F - \frac {\dot C} {A}\Biggr) - \frac {B}{A^2} (A \ddot C - \dot A \dot C). \label{Cdp1}
\end{equation}
Substituting for $\omega$ in Eq. (\ref{prdot}) from Eq. (\ref{omega}) gives, using Eq.  (\ref{f1})

\begin{equation}
\dot {C}^{\prime}=  \frac {\dot B} {B} C^{\prime} - \biggl[ {-{\frac{1}{{2C^2}}}+{\frac{
\dot{C}}{{AC}}}{\frac{C^{\prime}}{{BC}}}+{\frac{1}{2}}\biggl({\frac{
C^{\prime}}{{BC}}}\biggr)^2} + {\frac {\mu_{\Lambda}} {2}}\biggr]\; ABC.  \label{Cdp2}
\end{equation}
From equations  (\ref{Cdp1}) and (\ref{Cdp2}) we obtain

\begin{equation}
\frac {\dot C^3}{A^2} + \frac {2 C \dot C \ddot C} {A^2} - \frac {2 C \dot C^2 \dot A}{A^3} - \dot C (F^2 -1) - \mu_{\Lambda} C^2 \dot C  = 0 . \label{Cdp3}
\end{equation}
Eq. (\ref{Cdp3}) can be rewritten as

\begin{equation}
\frac {\partial}{\partial w} \Bigg[ \frac {C \dot C^2} {A^2} -  C (F^2 -1) - \frac {\mu_{\Lambda} C^3}{3}\Biggr]  = 0 .  \label{Cdp4}
\end{equation}
Hence, the expression within the square brackets depends only on the radial null coordinate $y$ and we define

\begin{equation}
 2 M(y) \equiv  \frac {C \dot C^2} {A^2} -  C (F^2 -1) - \frac {\mu_{\Lambda} C^3}{3} ). \label{Mdef}
\end{equation}
From Eq. (\ref{Mdef}) we find that 
\begin{equation}
 \frac {\dot C} {A}  = \pm \Biggl[ \frac {2 M(y)}{C} + (F^2 -1) + \frac {\mu_{\Lambda} C^2}{3}\Biggr] ^{1/2} .  \label{Cdp5}
\end{equation}
Substitution of Eq. (\ref{Cdp5}) into Eq. (\ref{f1}) gives

\begin{equation}
 \frac {C^{\prime}} {B} = F - \frac {\dot C}{A} = F \mp  \Biggl[ \frac {2 M(y)}{C} + (F^2 -1) +  \frac {\mu_{\Lambda} C^2}{3}\Biggr] ^{1/2} . \label{Cdp6}
\end{equation}
Now, for $y = y_{max}$, $C^{\prime}=0$. Therefore,
 
\begin{equation} 
6M_{max} + \mu_{\Lambda} C^3_{max} - 3 C_{max} = 0 .\label{hellabyeq}
\end{equation}
Eq. (\ref{hellabyeq}) originally obtained by Hellaby  \cite{Hellaby} in the 3+1 framework has to be considered a fundamental relation in Observational Cosmology, since it enables, from $C(w_0, z_{max})$ and $z_{max}$ measurements,  the determination of the unknown constant $\mu_{\Lambda}$ (see below).

Returning now to the main thread of our solution scheme, we
substitute for $C^{\prime}/B$ in Eq. (\ref{omega}) from Eq. (\ref{f1}) and use of equations (\ref{omegdef})
and (\ref{Mdef}) gives 

\begin{equation} 
M(y) = - \Biggl( \omega_0(y) +\frac{\mu_{\Lambda}}{6}\Biggr) C_0^3(y) = \frac {1}{2}\int_0^y{\mu_{m_0}(\tilde y)\;{B(w_0,\tilde y)}\;{C^{2}(w_0,\tilde y)}\;F(\tilde y)\;d\tilde y}.  \label{Mdef2}
\end{equation} 

From equations  (\ref{N}) and (\ref{Mdef2}) we find that the mass parameter $M(y)$ is related to number counts $N(y)$ as follows:

 \begin{equation}
M(y) = \frac {1}{8 \pi} \int_0^y mN^{\prime}(\tilde y) F(\tilde y) d\tilde y  =  \frac {1}{8 \pi} \int_0^y \bar M^{\prime}(\tilde y) F(\tilde y) d\tilde y . \label{MN} 
 \end{equation}
 where, $ \bar M(y) = m N(y)$ is  the total mass summed over the whole sky by a central observer out to a null radial distance $y$.

Stoeger {\it et al} \cite{OC III} and Maartens {\it et al} \cite{InhomUniv}
have shown that equations (\ref{f1}) and (\ref{f2}) can be transformed into
equations for $A$ and $B$, 
thus reducing the problem to determining $C$:
\begin{eqnarray}
&&A = {\frac{{\dot C} }{[F^2 -1 +2M/C+ (\mu_{\Lambda}/3) C^2]^{1/2}}}  \label{aeq} \\
&&B = {\frac{C^{\prime} }{F \pm [F^2 -1 +2M/C + (\mu_{\Lambda}/3) C^2]^{1/2}}}.  \label{beq}
\end{eqnarray}
\end{widetext}
The LTB form of the exact solution (Lema\^{\i}tre \cite{Lemaitre}, Tolman \cite{Tolman}, Bondi \cite{Bondi}; and cf. Humphreys \cite{HT} and 
references therein) is
obtained by integration of (\ref{aeq}) along the matter flow
$y= $constant using (\ref{tw})
\begin{equation}
t-T(y)=\int \frac {dC}{[F^2 -1 + 2M/C + (\mu_{\Lambda}/3) C^2]^{1/2}},  \label{bs}
\end{equation}
where $T(y)$ is arbitrary, and we identify
\begin{equation}
F^2=1-kf^2,\;\;\;\;k=0,\pm 1.  \label{f}
\end{equation}
Here $f=f(y)$ is a function commonly used in describing LTB 
models in the 3 + 1 coordinates \cite{Bonnor}.  Hellaby and Alfadeel \cite{helalf} have 
made clear in their paper that they were unable to find a time evolution equation for $A(w,y)$ without
resort to integrating over the $3+1$, $(t,y)$ domain first. It is precisely Eq. (\ref{bs}) above that is used to initiate
their indirect procedure to find $A(w,y)$.  In the next section we find a time evolution
equation for $(A(w,y)$ and a path towards full integration off our past light cone which
avoids this detour.

\section{\label{sec:nonflat}The General Solution - Time evolution off our Light Cone}

In this section we describe in detail  the general integration procedure that is applicable 
to all inhomogeneous spherically symmetric universe models  - that is the only constraint.  
We do not know whether the universe is homogeneous  or not. But the data gives us 
redshifts $z$, observer area distances (angular-diameter distances) $r_0(z)$, 
``mass source densities'' $M_0(z)$, and the angular-distance maximum
$C_{max}(w_0, z)$ at $z_{max}$. It is important to specify the latter, because,
as we have already emphasized, without them, we do not have enough information
to determine all the parameters of the space-time in the $\Lambda \neq 0$ case.
For instance, although we can determine $C(w_0, z)$ with good precision
(by obtaining luminosity distances $d_L$ and employing the reciprocity theorem,
equation  (\ref{recth})) out to relatively high redshifts, at present we do not yet have
reliable data deep enough to determine $C_{max}$ and $z_{max}$. But this has
just recently become possible with precise space-telescope distance
measurements for supernovae Ia. \\

Mustapha, {\it et al} \cite{must} have shown that a $\Lambda = 0$ LTB model can fit
any reasonable set of redshifts, observer-area-distance and galaxy-number-counts (or equivalently
our $M_0(z)$) data. However, as Hellaby \cite{Hellaby} first recognized, and as we
have been insisting here (see also Krasi\'{n}ski, {\it et al} \cite{kras}, and references therein),
$C_{max}(w_0, z_{max})$ data place an additional constraint on solutions, which may require a 
nonzero $\Lambda$. This is very important to determine. Even allowing for large-scale
inhomogeneities, is there definite evidence for nonzero vacuum energy, or some other
form of dark energy?  

In pursuing the general integration with these data, we use the framework
and the intermediate results we have presented in Section \ref{sec:fieqs}. Obviously, one
of the key steps we must take now is the determination of the ``potential''
$F(y)$, given by Eq. (\ref{f1}). This was done in a similar way for 
$\Lambda = 0$ by Ara\'{u}jo and Stoeger  \cite{AS}, as indicated above. This means
we need to determine $C^{\prime}(w_0, y)$ and $\dot{C}(w_0, y)$, which we
now write as $C_0^{\prime}$ and $\dot{C}_0$, respectively. We also need
$A(w_0, y).$ We remember, too, that at on $w = w_0$ we have chosen
$B(w_0, y) = A(w_0, y)$, which we have the freedom to do. \\

Clearly, $C_0^{\prime}$ can be determined from the $r_0(z) \equiv C(w_0,z)$
data, through fitting, along with the solution of the null Raychaudhuri Eq. (\ref{nr}) to 
obtain $z = z(y)$  (Stoeger {\it {et al}}  \cite{OC III}).  $A(w_0, y)$,
too, is obtained from redshift data along with this same $z(y)$ result. We pause
here to mention that this latter result together with Eq. (\ref{z}) allows us to determine
$A(w_0, y)$ and its value on ${\cal C}$, $A(w_0, 0)$. The unknown constant $A(w_0,0)$ in 
Eq. (\ref{z}) and in the solution of Eq. (\ref{nr}) cancel when we equate the two results, 
allowing us obtain an explicit expression for $A(w_0, y)$. Setting $y = 0$, we then 
find $A(w_0, 0)$ precisely. In a way Eq. (\ref{z}) provides us with a central condition
at $w = w_0$.
  
$\dot{C}_0$ is somewhat more difficult to determine. But the procedure is
straight-forward.\\

We determine $\dot{C}_0$ by solving Eq. (\ref{prdot}) for it on
$w = w_0$. Using equations (\ref{ab}) and (\ref{coneq}), we can write this now as:
\begin{equation}
\frac{\dot{C}_0^{\prime}(y)}{C_0(y)}= -\frac{A_0^{\prime}(y)C_0^{\prime}(y)}
{A_0(y)C_0(y)} - A_0^2(y)(\omega_0(y) + \mu_{\Lambda}/2). \label{cdpz}
\end{equation}
But, from Eq. (\ref{omega}) we can write $\omega_0(y)$ in terms of $C_0(y)$,
$C_0^{\prime}(y)$, and $\dot{C}_0(y)$. So Eq. (\ref{cdpz}) becomes:
\begin{widetext}
\begin{equation}
\dot{C}_0^{\prime}(y) + \frac{C_0^{\prime}(y)\dot{C}_0(y)}{C_0(y)} =
\frac{A_0^2(y)}{2C_0(y)} -\frac{A_0^{\prime}(y)}{A_0(y)}C_0^{\prime}(y) -
\frac{(C_0^{\prime}(y))^2}{2C_0(y)} - \frac{A_0^2(y)C_0(y)}{2}\mu_{\Lambda}.
\label{cdpzfin}
\end{equation}
\end{widetext}
This is a linear differential equation for $\dot{C}_0(y)$, where from data
we know everything on our past light cone, $w = w_0$, (once the null
Raychaudhuri Eq. (\ref{nr}) has been solved) except $\dot{C}_0(y)$ itself and
$\mu_{\Lambda}$, which is a constant that can be carried along and determined
subsequently from $C(w_0, z_{max})$ and $z_{max}$ measurements (see below). 
Thus, we can easily solve Eq. (\ref{cdpzfin}) for $\dot{C}_0(y)$, which
will also depend on the unknown constant $\mu_{\Lambda}$. Its general solution is given by:

\begin{equation}
\dot C_0(y) = \frac{K}{C_0(y)}+\frac {1}{2C_0(y)} \int_0^y \Biggl( A_0^2(\tilde y) - \frac {2 A_0^{\prime}(\tilde y) C_0^{\prime}(\tilde y)C_0(\tilde y)}{A_0} -(C_0^{\prime}(\tilde y))^2-A_0^2(\tilde y)C_0^2(\tilde y)\mu_{\Lambda} \Biggr) d\tilde y.
\end{equation}
where K is an integration constant to be determined by a boundary condition that, in our case, is a central condition for $\dot C(w,y)$. From Eq. (\ref{cent}) we can easily find that 

 \begin{equation}
{\rm as}\;\;y\rightarrow 0:\;\;\; \dot C(w,y)\rightarrow \dot B(w,0)y = 0.  \label{centC0} 
\end{equation}
Since from Eq.  (\ref{cent})  $C_0(y) \rightarrow 0$ as $y \rightarrow 0$ we find that the constant $K$ has to be set to zero if Eq.  (\ref{centC0}) is to be satisfied.
This procedure enables us to determine $F(y)$, that obviously also depends on   $\mu_{\Lambda}$.  Our next
step is to insert this result along with  $N^{\prime}(y)$ into  equation  (\ref{MN}) to determine the mass function $M(y)$. 
Next we evaluate the mass function $M(y)$ at $y_{max}$ and plug the result into Eq. (\ref{hellabyeq}) which becomes an algebraic equation for  $\mu_{\Lambda}$. With this determination of $\mu_{\Lambda}$, we know $\dot{C}_0(y)$ completely,
and can now determine $F(y)$ from Eq. (\ref{f1}). From here on, we can now follow the solution off $w = w_0$ for all $w$ as
we demonstrate in detail below.

We now turn to deriving an equation for $\dot{A}(w_0, y)$, which leads to finding 
subsequent equations for all higher derivatives of the metric variable on our
past light cone. This leads to our key result.

Equation  (\ref{bdd}) can be rewritten as

\begin{equation}
\frac {A}{B} \frac {\partial}{\partial w} \Biggl(\frac{\dot{B}} {A}\Biggr) = -2\omega \;A^2-{
\frac 12}\mu_{m} \;A^2 . \label{bddm}
\end{equation}

Substitution of equation  (\ref{coneq}) into  (\ref{bddm}) yields

\begin{equation}
\frac {\partial}{\partial w}\Biggl( \frac{A^{\prime}} {A}\Biggl) =  \frac {\partial}{\partial w} \Biggl[\frac{\partial} {\partial y} (\ln A)\Biggr] = \frac {\partial}{\partial y} \Biggl[\frac{\partial} {\partial w} (\ln A)\Biggr] = \frac {\partial}{\partial y}\Biggl( \frac{\dot{A}} {A}\Biggr) = 2\omega \;AB + {\frac 12}\mu_{m} \;AB \label{Adot}
\end{equation}
where we used the fact that $w$ and $y$ are independent coordinates. 

On our past light cone $(w=w_0)$ the above equation reads

\begin{equation}
\frac {\partial}{\partial y} \Biggl[\frac{\dot{A}(w_0,y)} {A(w_0,y)}\Biggr] = \Biggl[2\omega_{0}(y) +{
\frac 12}\mu_{m_{0}}(y)\Biggl]\;A^2(w_0,y). \label{Aplc}
\end{equation}

Therefore, its general solution is

\begin{equation}
\dot{A}_{0}(y) = A_{0}(y)\Biggl\{ \int_0^y\Biggl[2\omega_{0}(\tilde y) +{
\frac 12}\mu_{m_{0}}(\tilde y)\Biggl]{\;A_{0}}^2(\tilde y)\;d{\tilde y} + C_1\Biggr\}. \label{Adplc}
\end{equation}
where we have written $\dot{A}_0(y)$ and  $A_0(y)$  for  $\dot{A}(w_0,y)$
and $A(w_0,y)$ respectively.  $C_1$ is an integration constant that one would expect, in principle, to be determined by a central condition for $\dot A_0(y)$, namely, $\dot A_0(0)$. (As mentioned above, we have already determined $A_0(0)$, the first term of our Taylor series for $A(w, y)$.) However, this procedure does not fix $C_1$, since the central condition for $A(w,y)$ only specifies that it must not go to zero as y goes to zero.
Therefore, we have to investigate if use of the other central conditions provides a way of fixing $C_1$.
 
It is important to note that it is this equation which enables us to stay within the OC formalism during the integration, providing a much more direct and streamlined alternative to the Hellaby and Alfadeel \cite{helalf} scheme.

We observe that $\omega_{0} (y)$ is fully determined at this stage since
it can be obtained from the first equality in Eq.  (\ref{Mdef2}) as

\begin{equation}
\omega_{0} (y) = - \frac {M(y)} {C^3_0} - \frac {\mu_{\Lambda}} {6} \label{omegMmu}
\end{equation}

Hence, we have shown that the data on our past light cone determines $\dot{A}_{0}(y)$,  except for the constant $C_1$ which still remains undetermined at this stage. Since we know $\dot{A}_{0}(y)$ from the data,  equations  (\ref{Cdd}) and  (\ref{bdd}) evaluated on our past light cone become algebraic equations for  $\ddot{C}_{0}(y)$  and $\ddot{B}_{0}(y)$, respectively

\begin{eqnarray}
\ddot{C}_{0}(y) &=&  \frac{\dot{C}_0(y){\dot{A}_0}(y)}{A_0(y)}+\Biggl[\omega_{0}(y) +{
\frac 12}\mu_{m_{0}}(y)\Biggl]{\;A_{0}}^2(y)C_0(y) \label{Cddpnc} \\
\ddot{B}_{0}(y) &=& - \frac{A_0^{\prime}(y){\dot{A}_0}(y)}{A_0(y)}-\Biggl[2\omega_{0}(y) +{
\frac 12}\mu_{m_{0}}(y)\Biggl]{\;A_{0}}^2(y)B_0(y)  \label{Bddpnc}
\end{eqnarray}
where in the later we have used Eq. (\ref{coneq}). Note that $\ddot{C}_{0}(y)$ is completely determined at this stage except for its dependency on $\dot{A}_{0}(y)$ which carries its $C_1$ dependency  as we have explained above. However, $\ddot{C}_{0}(y)$ must satisfy the central condition for $ \ddot{C}_{0}(w,y)$, that 
from Eq. (\ref{cent}) is

 \begin{equation}
{\rm as}\;\;y\rightarrow 0:\;\;\; \ddot C(w,y)\rightarrow \ddot B(w,0)y = 0,  \label{centC00} 
\end{equation}
which, applied on $w=w_0$, fixes the constant $C_1$, and we are assured that $\ddot{C}_{0}(y)$ is completely determined by the data on our past light cone.

Our next step is to differentiate Eq.  (\ref{bdd}) with respect to $w$, that is,

\begin{equation}
\frac{\partial^3 B}{\partial w^3} - \frac{\ddot A \dot B}{A}=\frac{\ddot B \dot A}{A}-\frac{\dot B  \dot A^2}{A^2} -\frac{\partial}{\partial w}\Biggl[ \Biggl(2\omega +{\frac 12}\mu_{m} \Biggl)\;BA^2\Biggl]  \label{Bddd}
\end{equation}

Substitution of  Eq.  (\ref{coneq}) and its second time derivative  into the L.H.S. of  Eq. (\ref{Bddd}) gives

\begin{eqnarray}
\frac{\ddot A A^\prime}{A} - \ddot A^\prime&=&\frac{\ddot B \dot A}{A}-\frac{\dot B  \dot A^2}{A^2} -\frac{\partial}{\partial w}\Biggl[ \Biggl(2\omega +{\frac 12}\mu_{m} \Biggl)\;BA^2\Biggl] \nonumber \\
\frac{\partial}{\partial y}\Biggl(\frac {\ddot A}{A}\Biggl)&=& -\frac{\ddot B \dot A}{A^2}+\frac{\dot B  \dot A^2}{A^3} +\frac{1}{A}\frac{\partial}{\partial w}\Biggl[ \Biggl(2\omega +{\frac 12}\mu_{m} \Biggl)\;BA^2\Biggl] \label{Addp}
\end{eqnarray}

Evaluating Eq.  (\ref{Addp})  on our past light cone gives

\begin{eqnarray}
 \frac{\partial}{\partial y}\Biggl[\frac {\ddot A_0(y)}{A_0(y)}\Biggl]&=& -\frac{\ddot B_0(y) \dot A_0(y)}{A^2_0(y)}+\frac{\dot B_0(y)  \dot A^2_0(y)}{A^3_0(y)}    \nonumber\\
 &  & \hfill{\qquad} +\frac{1}{A_0(y)} \Biggl\{ \frac{\partial}{\partial w}\Biggl[ \Biggl(2\omega +{\frac 12}\mu_{m} \Biggl)\;BA^2\Biggl]\Biggl\}_0  \label{Addpplc}
 \end{eqnarray}

Therefore, its general solution is

\begin{eqnarray}
\ddot A_0(y)&=&-A_0(y) \Biggl\{\int^y_0 \frac{\ddot B_0(\tilde y) \dot A_0(\tilde y)}{A^2_0(\tilde y)} -\frac{\dot B_0(\tilde y)  \dot A^2_0(\tilde y)}{A^3_0(\tilde y)}  \nonumber\\
&  & \hfill{\qquad}  -\frac{1}{A_0(\tilde y)} \biggl\{\frac{\partial}{\partial w}\biggl[ \biggl(2\omega +{\frac 12}\mu_{m} \biggl)\;BA^2\biggl]\biggl\}_0 d \tilde y + C_2 \Biggr\}, \label{Addplc}
\end{eqnarray}
where, using the same reasoning as above, $C_2$ is an integration constant to be determined by the central condition for $\partial^3_wC(w_0,y)$, similar to Eq. (\ref{centC00}),  which is found as the solution of an algebraic equation once we differentiate Eq. (\ref{Cdd}) with respect to $w$ and substitute for the previously determined quantities.

It is important to note that all quantities on the R.H.S. of the above equation are obtainable either directly from the data or from the algorithmic steps in the procedure we are
describing here (Appendix). Therefore, we have shown that  we can obtain  $\ddot A_0(y)$ from the data. It is clear now that repetition of this procedure will give us all time derivatives
of A, B and C on our past light cone, which means that $A(w,y)$ $B(w,y)$ and $C(w,y)$ are completely determined by data on our past light cone, and calculable as Taylor series. 

It is clear from the above procedure that each step begins by finding the successive time derivatives of the metric function $A(w,y)$ on our past light cone, $\partial^n_wA(w_0,y)$. For its complete determination
one must apply the corresponding central condition, that is, one must specify $\partial^n_wA(w_0,0)$, which is done through other higher-derivative central
conditions of the form given in Eq. (\ref{centC00}).
Hence, it is shown that $A(w,0)$ is completely determined by the central conditions, thus proving the conjecture discussed in the introduction that the fulfillment of assumptions $(1)$ and $(3)$ removes the freedom of rescaling the time coordinate $w$ and completely determines $A(w,y)$. 
  
\section{\label{sec:concl}Conclusion}

We have summarized the essential details of previous work showing how to construct all spherically symmetric,
inhomogeneous cosmological (LTB) models in observational coordinates from cosmological data on our past light cone,
allowing for a nonzero cosmological constant (vacuum energy). In doing so we provide a new rigorous demonstration of how 
such data fully determines the time evolution of all the metric components, and a Taylor series algorithm for
determining those solutions. This enables us to move the solution we obtain from data on the light cone 
off it in a well-defined and straightforward way. It is essential for these to have data giving the maximum of 
the observer area (angular-diameter) distance, $C_0(w_0, z_{\max})$,
and the redshift $z_{max}$ at which that occurs. This enables the determination
of the vacuum-energy density $\mu_{\Lambda}$, which would otherwise remain
undetermined. That is, do we need $\Lambda \neq 0$ to adequately fit observational 
data, if we do not assume that the Universe is FLRW? Using this broader theoretical 
framework will enable us eventually to answer this important question, as well as to determine
more securely how close or far the universe on large scales is from being FLRW. \\

\section{\label{sec:Appen}Appendix}

The Algorithm:

(i) Solve the null Raychaudhuri Eq. (\ref{nr}) to obtain $z = z(y)$.

(ii) Determine $C_0^{\prime}$ from the $r_0(z) \equiv C(w_0,z)$
data, through fitting, along with the solution of the null Raychaudhuri equation.

(iii) Determine $\dot{C}_0$ by solving Eq. (\ref{prdot}) for it on $w = w_0$.  Note that at this stage $\dot{C}_0$ depends on  $\mu_{\Lambda}$.

(iv) Determine $F(y)$ from Eq. (\ref{f1}) which also depends on  $\mu_{\Lambda}$.

(v) Determine the mass function $M(y)$ from equation (\ref{MN}) using  $F(y)$  along with $N^{\prime}(y)$ 
that is given from data.

(vi) Evaluate the mass function $M(y)$ at $y_{max}$.

(vii) Determine $\mu_{\Lambda}$ from Eq. (\ref{hellabyeq}).

(viii) Knowing $\mu_{\Lambda}$, we determine both $\dot{C}_0(y)$ and $F(y)$ completely. 

(ix) Determine $\omega_{0} (y)$ from equation (\ref{omegMmu}). From there on, we can follow the solution off $w = w_0$ for all $w$.

(x) Determine $\dot{A}_0(y)$ using the procedure described between equations (\ref{bddm}) and (\ref{Adplc})
applied to equation (\ref{bdd}).

(xi) Determine both $\ddot{C}_{0}(y)$  and $\ddot{B}_{0}(y)$ algebraically from equations (\ref{Cdd}) and  (\ref{bdd}), evaluated on our past light cone, respectively.

(xii) Differentiate equations  (\ref{bdd})  (\ref{Cdd}) with respect to $w$. 

(xiii) Go back through steps (x) to (xii) to find at each run $\partial^n_wA(w_0,y)$ , $\partial^n_wB(w_0,y)$ and
 $\partial^n_wC(w_0,y)$ respectively.

\end{document}